\documentclass[conference]{IEEEtran}
\renewcommand\IEEEkeywordsname{Keywords}
\IEEEoverridecommandlockouts
\usepackage{cite}
\usepackage[overload]{textcase}
\usepackage{enumitem,amsmath,amssymb,amsfonts}
\usepackage{algorithm}
\usepackage{algorithmic}
\usepackage{graphicx}
 \graphicspath{ {Figures/} }
\usepackage{textcomp}
\usepackage{xcolor}
\def\BibTeX{{\rm B\kern-.05em{\sc i\kern-.025em b}\kern-.08em
    T\kern-.1667em\lower.7ex\hbox{E}\kern-.125emX}}

\begin{document}

\title{Blockchain-based Platform for Secure Sharing and Validation of Vaccination Certificates   

{\footnotesize \textsuperscript{} }
}

\author{\IEEEauthorblockN{Mwrwan Abubakar, Pádraig McCarron, Zakwan Jaroucheh, Ahmed Al-Dubai  William J Buchanan}
\IEEEauthorblockA{\textit{Blockpass ID Lab} \\
\textit{Edinburgh Napier University}\\
Edinburgh, UK}}

% \author{\uppercase{OMITTED FOR SUBMISSION }\IEEEauthorblockA{}}

\maketitle
\begin{abstract}
The COVID-19 pandemic has recently emerged as a worldwide health emergency that necessitates coordinated international measures. To contain the virus’s spread, governments and health organisations raced to develop vaccines that would lower Covid-19 morbidity, relieve pressure on healthcare systems, and allow economies to open. As a way forward after the COVID-19 vaccination, the Vaccination certificate has been adopted to help the authorities formulate policies by controlling cross-border travelling. To resolve significant privacy concerns and remove the need for relying on third parties to maintain trust and control the user’s data, in this paper, we leverage blockchain technologies in developing a secure and verifiable vaccination certificate. Our approach has the advantage of utilising a hybrid architecture that implements different advanced technologies, such as smart contracts, interPlanetary File System (IPFS), and Self-sovereign Identity (SSI). We will rely on verifiable credentials paired with smart contracts to implement on-chain access control decisions and provide on-chain verification and validation of the user and issuer DIDs. The usability of this approach was further analysed, particularly concerning performance and security. Our analysis proved that our approach satisfies vaccination certificate security requirements.

\end{abstract}

\begin{IEEEkeywords}
COVID-19, Blockchain technology, smart contract, SSI, IPFS. 
\end{IEEEkeywords}

\section{Introduction}
The shared epidemiological information system was originally established in the 1920s by the newly formed League of Nations Health Organization in response to the 1918 pandemic outbreak. It utilised a novel communication technology at the time, wireless telegraphy, to share information about cases between moving ships on the sea and land before the cases arrived, thereby halting the disease's spread \cite{1}. Despite its long-forgotten history, this system can teach us important lessons about how international collaboration can use emerging technologies to tackle the recently discovered coronavirus-2019 (COVID-19). Due to its extreme contagiousness, COVID-19 has had an unprecedented worldwide impact. It has affected human health as well as the global economy due to harsh countermeasures such as lockdown implemented by governments around the world. The ability of researchers to harness science and new technology to develop treatments or exchange information about diseases has never been stronger than it is now. The capacity to quickly communicate information has been critical in the fight against COVID-19. As the globe mobilises to combat the COVID-19 epidemic, Information Technology (IT) has shown to be extremely beneficial in stopping the spread of the virus. To combat the COVID-19 epidemic, World Health Organization (WHO) has received enormous pro-bono assistance from technological businesses. On April 2, 2020, 40 of the world's leading digital technology professionals convened in a virtual roundtable to further the WHO's joint response to COVID-19 \cite{2}. Additionally, various open-source IT platforms have emerged, facilitating data exchange and enabling a better coordinated worldwide response to the infection. For instance, contact tracing provided a systematic approach to curb infectious disease transmission. Modes of delivering healthcare services have also been reshaped by technology to break the transmission cycle of COVID-19. For instance, several countries have enhanced their use of robots to disinfect areas, transport medical samples, automate testing and promote public safety \cite{3}. Drones have also been used to carry crucial medical supplies to rural areas \cite{4}. However, all sectors must act to ensure compliance with baseline COVID-19 mitigations and further targeted, and proportionate steps must be considered to further decrease risk. Vaccine certification is one example of such a measure. Vaccination certificates aid in the prevention and control of COVID-19 spread by enabling the authorities and governments to develop rules that permit cross-border travel for those who possess this certificate. Unlike traditional paper vaccination certificates, electronic documentation will increase interoperability and accountability by reducing data inconsistency across several sources, such as medical centres or governments. Thus, it would be more credible to follow pandemic progress across many countries.

\subsection{Problem statement}
While the realisation of vaccination certificates represents the main objective to prevent and control the spread of COVID-19 and helps locked down countries reopen their economies, the vaccination certificate's security is also associated with significant challenges. It will undoubtedly be a challenge to keep this information accurate and secure and work across different systems that will consume this data. According to \cite{5}, forged Covid-19 documentation is now being openly sold on dark web in 29 countries with prices ranging from US\$85-US\$200. In addition, we have seen how cyber-attackers have targeted Ireland's healthcare system early this year in May, which resulted in the Department of Health shutting down their IT system and causing substantial cancellations to outpatient services \cite{6}. In another instance, an independent researcher discovered a significant flaw in the Services Australia COVID-19 digital vaccine certificate that allows an attacker to forge someone's vaccination status \cite{7}. So, therefore, the recent availability of vaccines for COVID-19 makes it urgently necessary to consider proper infrastructures for vaccination certificates. Issues with current solutions are that they are centralised and lack privacy.

\subsubsection{Centralisation} The legacy authentication and access control practices used to get access to critical COVID-19-related data are mostly centralised, necessitating the establishment of a centralised trust for proper operation. Authentication data is stored and managed by many service providers. Furthermore, the centralisation of effective security solutions, such as Public Key Infrastructure (PKI), would cause scalability issues due to the thousands of nodes connected \cite{8}. Furthermore, working on a centralised authority comes with its own set of vulnerabilities, such as a single point of failure. As a result, collaboration options among collaborating organisations may be limited.

\subsubsection{Lack Privacy} Issues with centralised trust is often solved by utilising third-party data backup providers. By relying on a third party to collect and analyse such data, the risk of exposure is effectively increased. As a result, digital vaccination certificates have cast doubt on the protection of certain persons' fundamental rights. As relying on third parties to manage the trust could enable the acquisition of personal information without the users' agreement, resulting in major data leakage risks. Additionally, individuals have limited or no control over their sensitive personal information. \\ 

To help achieve self-sovereignty and increase the adoption of the digital vaccination certificate, there is a need for more dependable solutions with privacy-preserving and user-centric access control. On the other hand, blockchain and decentralised approaches have tremendous opportunities in this context. The decentralisation of trust is increasingly becoming a dominant direction, creating potential opportunities to manage authentication and access control in a decentralised and autonomous manner.  Blockchain coupled with smart contract technology \cite{10} eliminates the reliance on central servers to manage the trust. All the connected entities on the blockchain network will have a copy of them, which provide an equal right to control all contract operations. In addition, blockchain provides immutability properties. This means that transactions recorded in the blockchain cannot be modified or altered, guaranteeing high system trustworthiness and integrity. Besides, access control using blockchain can achieve the transparency property with the ability to effectively solve the issue of data leakage, which can be caused by relying on a third party \cite{10}. 

Motivated by such advantages of blockchain and distributed ledger technology, in this paper, we aim to build a blockchain-based platform for sharing vaccination data to enable a secure and privacy-aware verification process. For this, we utilised the Ethereum blockchain system and smart contract technology to build a secure and scalable decentralised architecture for vaccination certificates creation and verification. The remainder of this paper are structured as follows. After introducing the topic in Section 1, we provide an overview of related work in Section 2 and summarise our main contribution. In section 3, we present a background on blockchain technology and the Self-Sovereign Identity (SSI) principles. Subsequently, we elaborate on our proposed solution in section 4 and describe the system design in section 5. Then, we present the implementation of our system in section 6. We further provide the performance and security analysis of our approach in section 7. Finally, we conclude the paper in Section 8.

\section{Related work and Our Contribution}
Since the onset of the COVID-19 pandemic, the scientific community has taken a keen interest in health data exchange and monitoring. A significant of current research focuses on developing a secure and privacy-preserving method for sharing and verifying vaccination certificates. The study presented in \cite{11} investigated the currently proposed blockchain solutions for mitigating the COVID-19 challenges. The authors identified that the most prevalent applications of blockchains for limiting COVID-19 implications are contact tracing and support for immunity and vaccination certificates. Additionally, a European Parliament official study \cite{12} names blockchains as one of 10 technologies to tackle COVID-19 and cites infection tracking and health data monitoring as two of the most prominent cases. For the realisation of this COVID-19 vaccination data-sharing platform, blockchain technology has been postulated in different related scenarios. For instance, the authors of \cite{13} proposed Verifiable Credentials (VC) for the vaccination certificates that can be stored directly on the user mobile phone. The data relating to vaccination can be stored in different organisations and can only be reviewed through a cryptographical mechanism. However, the presented work was just a conceptional proposal without providing proof of concept or evaluation. Another use of the VC was presented in \cite{14}, where the authors developed a prototype mobile phone app and the decentralised server architecture. The presented work relies on the Verifiable Credentials concepts. Alternatively, several research efforts were proposed blockchain and smart contracts for securing vaccination certificates. For instance, the authors of \cite{15} present the concept of digital health passports, which rely on a private blockchain that use the proof-of-authority consensus for registering and storing test results. Similarly, the work presented in \cite{16} proposes a blockchain-based platform for the secure sharing of COVID-19 or other disease vaccination certificates. The proposed architecture relies on the deployment of Hyperledger Fabric on an emulated network in EPIC. Another solution was proposed in \cite{17}, which suggests the use of the InterPlanetary File System (IPFS) to decentralise the storage of medical tests and travel history. The presented approach is also utilising the power of smart contracts provided by Ethereum. However, most of the proposed approaches presented a general high-level description of the architectures of the blockchain-based systems. Instead, we aim to present a proof of concept to demonstrate our approach. Compared to previous methodologies, our approach has the advantage of utilising a hybrid architecture that implements different advanced technologies, such as the smart contract, IPFS system, and SSI concepts.  

\subsection{Our Contribution}
Our main contributions in the paper can be summarised as follows:
\begin{itemize}
\item We propose a blockchain-based solution for the creation and validation of vaccination certificates.
\item We developed a Self-sovereign Identity (SSI) model for secure sharing and verifying vaccination certificates.
\item A smart contracts to implement on-chain access control decisions.  
\item A proof-of-concept implementation of the proposed solution along with performance and security
Analysis to verify the feasibility of our solution.

\end{itemize}

\section{Background on Blockchain Technology}
Blockchain is a peer-to-peer network that keeps track of transactions in a distributed, decentralised, and immutable manner \cite{18}. A node is any device that contributes to the network. A transaction is a data unit on the Blockchain, and a Block is a collection of transactions. The cryptographic hash of a Block in the distributed ledger is used to link it to a previously approved Block \cite{18}. Because there is no single point of failure, the decentralised element of blockchain technology ensures the data and transactions stored on the blockchain are secure. The blockchain's transaction and data record is transparent to all network participants, creating trust in the data's trustworthiness and availability. The consensus mechanism is a system that enables the blockchain's network nodes to reach an agreement on the ledger's present state. Consensus is at the core of Blockchain technology since it ensures the network's integrity and security. Different blockchain systems employ different consensus mechanisms, and as a result, their operation and execution are distinct from one another \cite{19}. Cryptography and hashing methods ensure the integrity and immutability of transactions on the blockchain network. Asymmetric cryptography is used in blockchain to confirm the validity and integrity of data. Cryptographic hashing is also employed to connect each block to its predecessor. Blockchains can be classified as public, private, or consortium based on their membership. Any user can become a member of public blockchains. There are no limitations on membership, and they are just pseudo-anonymous. Public blockchains include Bitcoin and Ethereum. While private blockchains are owned and maintained by a business or organisation. On the other hand, a group of organisations or a private community owns and operates consortium blockchains. Due to the limited number of participants, private and consortium blockchains execute transactions more quickly than their public equivalents \cite{20}. Hyperledger and Corda are two examples of permissioned blockchains.
 
\subsection{Smart contract}
This term was first used by Nick Szabo in 1994 \cite{21} to refer to a theoretical concept of computer software that can be used to digitally enforce and verify contract negotiation. In the context of blockchain, this phrase refers to self-executed software that is stored on a blockchain system and executes when certain circumstances are met. Without the requirement for a central authority, legal system, or external enforcement mechanism, smart contracts enable the execution of trustworthy transactions and agreements between disparate, anonymous parties. While blockchain technology has historically been connected with Bitcoin, smart contracts in Bitcoin were limited to exchanging values between users when certain requirements are met. In Ethereum, however, the smart contract is treated as an entity with a unique account and address. This account can interface with other contracts, sending and receiving digital currency, as well as storing data. To allow developers to design their own smart contracts, Ethereum replaced Bitcoin's complex language with a Turing complete language called solidity. They are often used to automate the implementation of an agreement so that all participants are immediately certain of the outcome, with no intermediary involvement or time lost. They can also automate a workflow, activating the next operation when certain circumstances are satisfied \cite{10}.

\subsection{Decentralised identity}
Identity management methods are used to determine user identities. Third-party apps or protocols, such as single-point services or identity providers, control and preserve the identity of most modern systems' users. However, the primary challenge with data security is trusting the people or organisations entrusted to assure protection. In this situation, the specified suppliers own the identity, not the legitimate proprietors. With the advancement of blockchain technology, the use of self-sovereign identities (SSI) \cite{22} has expanded, leveraging the benefits of decentralisation. Users get authority and ownership of their own identities through the usage of SSI, as well as other benefits such as decentralised control and privacy. To actualize SSI, two critical standards must be implemented: Verifiable Credentials (VCs) \cite{23} and Decentralized Identifiers (DIDs) \cite{24}. VCs play a critical role in permitting authenticated attribute disclosure and privacy-aware identification, whereas DIDs focus exclusively on cryptographic identification.

\section{The proposed solution}
This section explains the proposed blockchain-based platform for secure and user-centric vaccination certificates verification. Our approach utilises the Verifiable Credentials (VC) for cryptographically using different keys to encrypt and decrypt sensitive information with the goal of defining applications respecting the fundamental rights of citizens. To resolve significant privacy concerns and remove the need for relying on third parties to maintain trust and control the user’s personal data used for authentication, we will rely on the Ethereum blockchain, which allows us to use smart contracts. We rely on the smart contract to implement on-chain access control decisions and to provide on chain verification and validation of the user and issuers DIDs. The users can interact with the smart contract by issuing transactions signed by their private key. The hash of the used key is taken as the user’s address and will be used to associate the user with their vaccination certificate. 
However, storing vaccination certificates information in the blockchain would be very expensive in terms of the storage needed and the time it takes to write data to the blockchain. For this, we propose using the InterPlanetary File System (IPFS) \cite{25} to store vaccination certificates and medical tests in a decentralised manner. During the program’s preliminary phase, after the user gets vaccinated, the vaccination provider centre will create a VC, which will be encrypted and stored into the IPFS system. Only the hash of VC location in the IPFS will be sent to the smart contract and mapped to the user public address. 
After which, we implement a decentralised identity model, which will be tasked with managing the users’ identities. This will be accomplished by utilising self-sovereign identity (SSI). With our approach, we no longer need a central trusted authority. The application also allows users to store their vaccination information straight on their mobile devices, which can be accessed during the verification phase, which uses our decentralised web application to request VC from the user. Due to the need for two shots of vaccines is required to be effective. Therefore, vaccination certificates could be only issued when both shots are administered to the citizen. However, using our approach, each vaccination shot could be associated with a VC so that it can be used to prove that a citizen has already received the first shot. In addition, by recording test information, our technique will further provide a solution for those who have not been vaccinated but have lately tested negative. This would accommodate vaccine passports for travel where both vaccination and a negative test are required while also allowing the use of the same system where either vaccination or a negative test is allowed. Our blockchain-based approach will allow interoperability and accountability properties so that potential discrepancies among data from different sources, such as medical centres or governments, could be avoided. A simple lookup in the distributed ledger will help verify the authenticity and integrity of the users’ information. In addition, the use of self-sovereign identities and verifiable credentials minimises the use of personal data putting the needed data under the holder’s control, underpinned by good security design practices based on the blockchain.

\section{System Design}
This section presents a detailed explanation of the design of the proposed decentralised testing and vaccination certification systems. Our system design consists of three main entities: the smart contract, the IPFS system, and the Self-Sovereign Identity (SSI) model to create and validate Verifiable Credentials (VC). Figure 4 presents a high-level design of a blockchain-based system to create a secure verifiable vaccination and testing certificate. 
 
\subsection{Blockchain and smart contract}
The proposed model exploits the advantages of blockchain and smart contracts technologies in developing a secure verifiable vaccination and testing certificate. We use Ethereum-based blockchain on the proposed model to store information in a distributed manner while maintaining consistency. Using an Ethereum-based blockchain is because it provides an open-source, public, distributed computing platform featuring smart contract (scripting) functionality. The smart contract is used in our approach to interfacing with data stored on the blockchain. They also offer resilience by executing smart contract code across all blockchain nodes. The web3 JS library, which uses the RPC to interact with the smart contract, will be used to access all the services written in the smart contract. The smart contracts will be used in our system to help to store an immutable record of users’ policies and authorisation information, the registration of medical centres. The certification of these centres can be made by the health authority. The smart contract will also implement policies such as on-chain access control decisions. Our smart contract can be used in two different scenarios: it can be used to issue immunisation certificates, and it can also be used to manage testing results.

\subsection{IPFS distributed file storage}
The InterPlanetary File System (IPFS) is used for decentralised off-chain document storing. It would be exceedingly expensive to keep all of the documentation associated with COVID-19 immunisation certificates and testing results on-chain. As a result, it is critical to store this data in a decentralised and secure manner. IPFS storage is distributed and is open to the public. As a result, data saved on IPFS should be encrypted, and only authorised entities should be able to view unencrypted content. Therefore, we proposed encrypting the certificate files in our system architecture before uploading them to the IPFS servers. Only the authorised receiver can access the cleartext because the smart contract governs access to an encrypted certificate from the IPFS system.  

\subsection{Self-sovereign identity (SSI) model}
\subsubsection{Issuer}

\begin{figure*}
    \centering
    \includegraphics[width=0.9\linewidth]{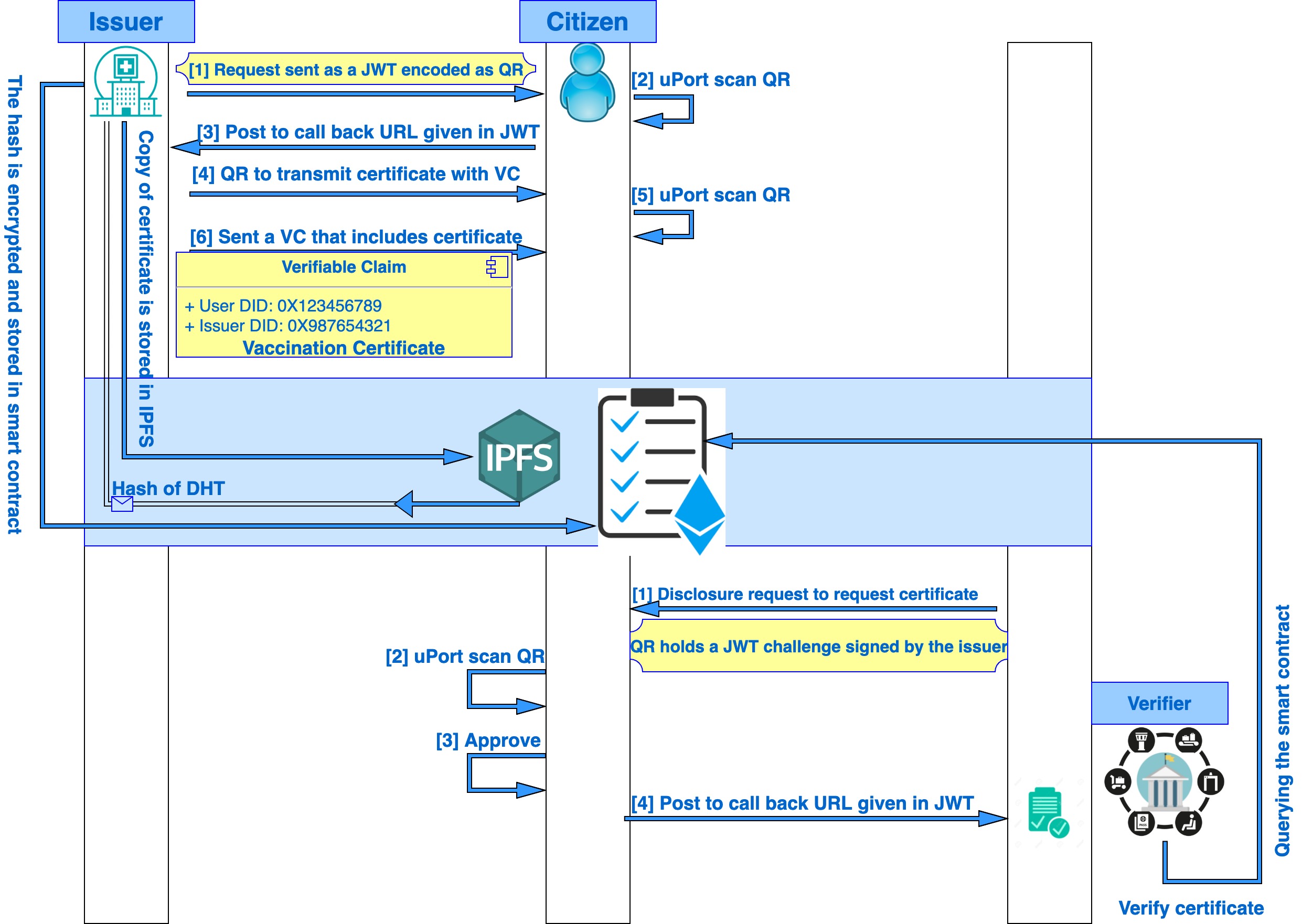}
    \caption{\textbf{The proposed System}}
    \label{fig:fig02}
\end{figure*}

Creating this certificate necessitates the establishment of an effective identity management architecture to associate citizen immunisation with their digital identity to ensure that the final certificate is scalable, verifiable, and preserving citizens’ privacy. This was accomplished in this case by utilising uPort’s decentralised identity capabilities \cite{26}, which enables issuing verifiable claims containing the vaccination certificate. By establishing an identity rooted in the Ethereum blockchain, uPort can leverage Ethereum's verification and security features. We employed DIDs along with VCs, in which the certificate’s issuer asserts vaccination information included in the VC in a way that can be verified by other verifiers that require proof of citizens’ vaccination. The system design is presented in figure 1. The citizen first needs to go to an authorised medical centre and present a valid identity document to get vaccinated and get a vaccination certificate issued. Because the vaccine is only effective after two doses. As a result, vaccination certificates could only be issued to citizens who had gotten both injections. In this case, our VC can represent that the citizen has received the first shot, so it can be used in administering the second one or can represent that the person is immune as they already received both shots. To ensure the VC’s legitimacy, the medical centre or a practitioner who performed the vaccine must sign it and include the issuer DID. Additionally, the citizen needs to produce a DID for embedding it in the VC using the uPort identification app. Additional information about the vaccine being given is contained in the VC, such as information regarding the specific doses of a particular vaccine. These bits of information may be used to certify and validate various vaccines. The VC will be encrypted and then sent to the IPFS, while only the IPFS hash is kept in the smart contract and mapped to the citizen’s decentralised identity, which representing the citizen public key pair. However, encryption of the certificate is out of the scope of this paper.

\subsubsection{Verifier}
Citizens who have gotten a vaccination certificate based on VC can use it to gain access to venues that need vaccination verification or can use it if travelling abroad. The VC is verified by a verifiable presentation in which the verifier seeks the vaccination certificate from the citizen. The VC request will be sent as a JWT encoded in a QR code, which the citizen Uport identity mobile app must scan. The verification service examines the medical centre's signature to ensure an authorised authority generates the credential. This can be achieved by querying the smart contract to match the issuer DID with the ones that are whitelisted. Furthermore, it validates the VC as well as the citizens' DID to confirm that it is the person linked with the credential supplied. These checks can be carried out by utilising the DIDs associated with the issuer (i.e., medical centre) and subject (i.e., the citizen DID) encoded in the VC. the verifier requests authentication from the user by presenting the challenge of uPort as QR code. The QR code is comprised of a JWT \cite{27} that is signed by the identity of our web application.

\subsubsection{Zero-Knowledge Proof (ZKP)}
Additionally, Covid-19 certificates would need to grant people a quick access into a venue based on a simple yes or no response. Our approach presented a zero-knowledge proof model in which the user can perform a verifiable presentation by displaying a portion of the credential's properties. Our solution enables selective publication of identity data while protecting users' privacy by utilising zero-knowledge proofs. This procedure may be advantageous for gaining admission to certain locations that demand merely confirmation of a person's vaccination status but not personal information.

\section{Implementation}
The use case developed in this paper involves a vaccination/testing centre who creates a VC that includes the citizen's vaccination certificate or a testing result. The citizens can receive their certificate directly into their mobile app, which can then be presented to a third party or a verifier to prove their vaccination status. The prototype is made up of five main components, including the smart contract, issuer, verifier, IPFS for decentralised data storage, and the decentralised VC recovery application.

\subsubsection{Smart contract implementation}
As a proof of concept, we implemented Ethereum smart contracts. The Ethereum network was chosen due to its capacity for smart contract deployment and the assistance that comes with its popularity. We chose Solidity \cite{28} to create our smart contracts, a turing complete language that enables the development of smart contracts in Ethereum blockchain systems. Then, we utilised the Ethereum-based integrated development environment (IDE) Remix \cite{29} to create, review, and deploy smart contracts over the Ethereum network. Additionally, the IDE includes a compiler for testing the functioning of smart contracts. Examples of these functions are presented below.  

\begin{algorithm}[h]
\caption{Only Owner}
\label{Algorithm 1}
\begin{algorithmic}[1]
\STATE modifier onlyOwner() \{.
\STATE require(msg.sender == owner);
\end{algorithmic}
\end{algorithm}

\begin{algorithm}[h]
\caption{Trusted Verifiers only}
\label{Algorithm 2}
\begin{algorithmic}[1]
\STATE modifier onlyWhitelisted() \{
\STATE require(isWhitelisted(msg.sender));
\end{algorithmic}
\end{algorithm}

\begin{algorithm}[h]
\caption{Centres addition}
\label{Algorithm 3}
\begin{algorithmic}[1]
\STATE function add\(address _address\) public onlyOwner \{
\STATE whitelist[\_address] = true;
\STATE emit AddedToWhitelis(\_address);
\end{algorithmic}
\end{algorithm}

\begin{algorithm}[h]
\caption{Centres Verification}
\label{Algorithm 4}
\begin{algorithmic}[1]
\STATE function isWhitelisted(address \_address) public view returns(bool) \{
\STATE return whitelist[\_address];	\}
\end{algorithmic}
\end{algorithm}

In addition, we used the web3.js Ethereum JavaScript API to interface with an Ethereum node running on Infura \cite{30} to facilitate communication between our application and the Ethereum network. The Ethereum-based Uport identity mobile application will be used as an Ethereum user's wallet. We used WebSockets as the transport to deliver a backup of vaccination certificates from our web application to an IPFS node.

\subsubsection{SSI implementation}
Using uPort identity functionality, we created the issuer and verifier interfaces as well as a decentralised identity architecture. Our web application is made up of a variety of resources. HTML templates, JavaScript files, CSS files, image files, and server-side implementation code are all examples of this. Implementing the SSI concepts in our proposed model has been associated with the use of Decentralized Identifiers (DID) and Verifiable Credentials (VC), which are being standardised by the World Wide Web Consortium (W3C). A DID is an identifier controlled by a DID subject that denotes a DID method and a specific identification for that method. Furthermore, A DID is also resolved to a DID document that defines the verification process.
To verify the VC, the verifier will ask for the citizen's vaccination certificate in a verifiable presentation. The VC request will be issued as a JWT embedded in a QR code. The uPort mobile app decodes the code and verifies the JWT's signature after it is scanned. In addition, the app points to the requested attributes and requests permission from the user to communicate the corresponding verified claims to our application. The user will be alerted of the request via their uPort app and given the option to decline or approve it using a fingerprint or mobile pin code. In response, the uPort identity mobile app creates a new JWT containing the verifiable claim and transfers it to the callback address included in the authentication challenge. Our decentralised web application will process the received JWT and verify the citizen’s signature.

\section{Evaluation}
This section assesses the proposed vaccine certificate platform and demonstrates its security and efficiency for application in real-world contexts, including performance study, financial costs, and its security properties. To assess our system's performance, we created the issuer and verifier web interfaces according to our planned design. Our testing configuration is based on a MacBook Air 2018 equipped with a 256GB SSD drive, 8GB RAM, and a 1.6 GHz dual-core Intel Core i5 processor. 

\subsection{Performance analysis}
\subsubsection{CPU utilisation} We measured storage and processor utilisation using the activity monitor application on our system. Our assessment discovered that even in the worst-case scenario, CPU use remains below 40\%. While the verification process, which involves requesting verifiable credentials (VC) from the citizen Uport app, consumes less than 20\% of the CPU power. This is due to the fact that it executes these operations off-chain. On the other hand, CPU utilisation for querying the smart contract to verify citizens' and issuers' DIDs is costing less than 3\% of available CPU power. This is because these operations are designed as view functions, which means they have no CPU overhead, delay, or cost because they just read the state of the blockchain without modifying it in any way. In general, we observed that our application has a low computational burden. Additionally, our technique consumes very little memory and storage space because the immunisation certificate is not saved locally by our application but is instead stored directly on the user's smartphone and/or on the IPFS.

\subsubsection{End to end delay} We used the internal time function of the MacBook to determine the total time required to issue and verify VC as shown in Fig. 2. On the x-axis, the number of concurrent users is shown. The y-axis depicts the execution time of a single request. The time required to request and verify VC from the user can be summarised as follows. The time required to request and verify VC from the user can be summarised as follows. The time required to construct the JWT authentication challenge delivered in the form of a QR code ranges from 1.5 seconds with one concurrent user to 7 seconds with 10 simultaneous users. While the time required for the complete verification process, including the creation and authentication of the JWT challenge and processing the response to receiving VC from the user, is approximately 5 seconds with a single concurrent user and the time required for 10 simultaneous users is around 22 seconds. Overall, the number of concurrent users increases the execution times significantly.

\begin{figure}[!ht]
    \centering
    \includegraphics[width=0.8\linewidth]{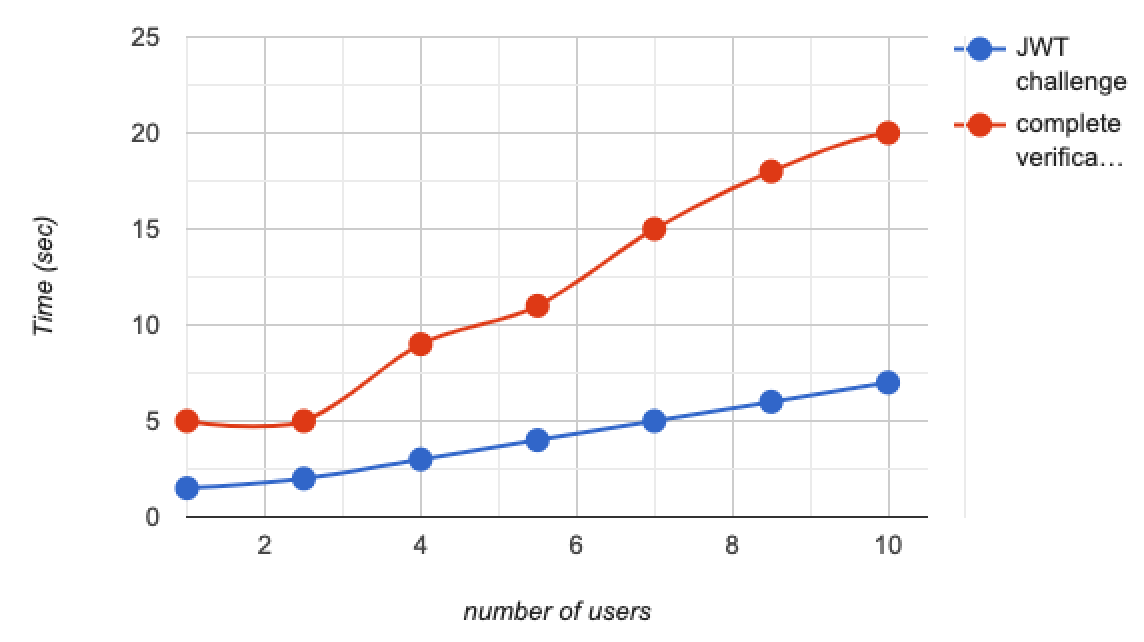}
    \caption{\textbf{End to end delay}}
    \label{fig:GeneralDiagram}
 \end{figure}
 
While the verification process has a lower execution time overhead, registration has a much greater one. This is because it needs to sign the transaction to be submitted to the smart contract. In the end, though, this is a problem that should be expected when using the open Ethereum network. This is because it takes an average of 13 seconds to complete the transactions. Overall, the number of concurrent users increases the execution times significantly.

\subsubsection{Transaction’s cost} 
\begin{figure}[!ht]
    \centering
    \includegraphics[width=0.8\linewidth]{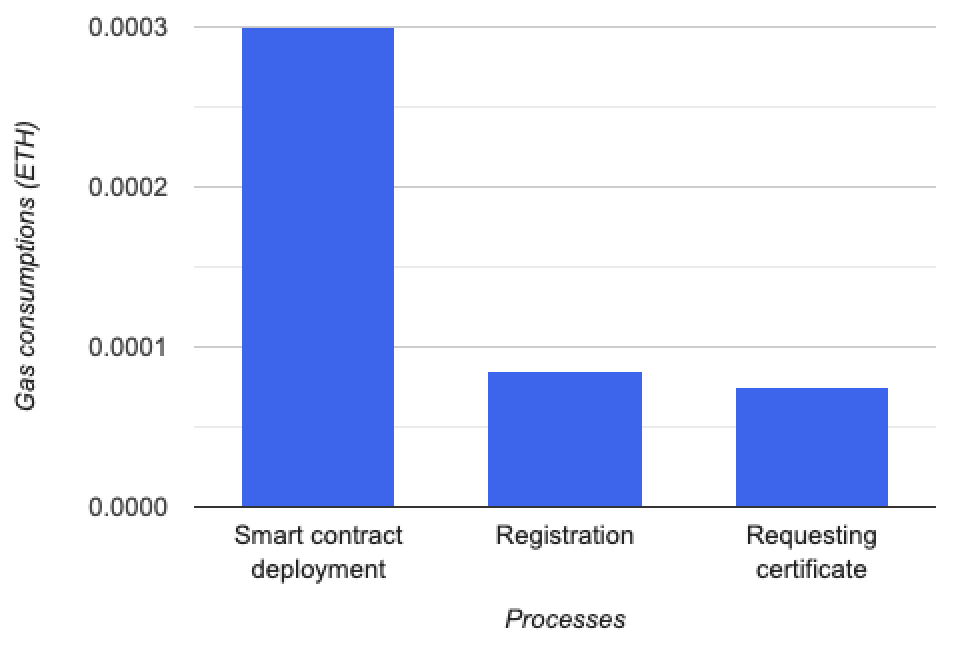}
    \caption{\textbf{Gas consumption}}
    \label{fig:GeneralDiagram}
 \end{figure}
 
 Additionally, each transaction and interaction with the smart contract on the Ethereum blockchain costs a specific amount of gas. As illustrated in Figure 3, the cost of deploying smart contracts is the highest. Nonetheless, this cost is variable, as it depends on various criteria, including transaction speed and required storage. However, it is vital to emphasise that our implementation is built on the Rinkeby testing environment, which is free to use because Ethereum offers this testing network with free Ethers.
 
\subsection{Security analysis}
Our solution complies with the three primary security standards. This is known as the CIA's security triad (Confidentiality, Integrity and Availability). Availability refers to the availability of services upon request. Integrity ensures that messages reach their intended recipients intact, while confidentiality ensures that the system is accessible only to those with authorization. We then conducted an analysis of our system's security in light of these security standards.

\subsubsection{Confidentiality}
The cryptographic technology used to establish a blockchain ensures that transactions recorded in the blockchain are exceedingly secure. Our solution will make use of the blockchain's asymmetric encryption technology to ensure that only authorised users have access to the system.

\subsubsection{Integrity}
It is incredibly impossible for anyone to edit or modify anything after it has been stored on the blockchain. As a result, because user access control rolls and challenges signed with the user's private key are immutable, the system does not allow them to be modified.

\subsubsection{Availability}
Our design ensures that data pertaining to the verification and authentication processes recorded on the blockchain is readily available at all times. The transaction data is replicated and updated on each node. It doesn't matter whether a node is removed from the network unintentionally, maliciously, or otherwise rendered unavailable; the network as a whole will continue to operate. As a result, our system will provide an extremely high level of availability.

\subsubsection{Auditing}
In addition, any node in the blockchain has the ability to approve any modifications made to any of the blockchain's linked blocks, regardless of where they occur. The ability to trace blockchain transactions and build high-quality security intelligence around those transactions allows organisations to assure the traceability of blockchain transactions while also enabling for the auditing and tracking of data changes.

\subsubsection{Security of our system}
Our evaluation revealed that our system would remain secure in the face of various threats. For instance, a common hazard to any website is attackers gaining access to a website where a user has an account. As a result, attackers may be able to access the user's password and other accounts that share the same password. In our solution we eliminated the need for a user password to gain access to a user's digital certificate in our system. Rather than that, we rely on verifiable credentials that refer to the DID assigned. Our strategy ensures an unwavering commitment to secrecy and integrity. It is supported by the issuer's DID and cryptographic proof in order to ensure the security of data transmission between users and our online application. However, while a lost or forgotten password may be readily changed in password-based systems, losing the private key results in asset loss in blockchain-based SSI systems. To mitigate these concerns, uPort provides a delegation mechanism into two unique contracts dubbed the controller contract and the proxy contract. This enables uPort users to reclaim their identities in the event of a private key loss. Furthermore, if the system is subjected to a man in the middle (MITM) attack, our proposed system will be immune because our model makes use of the cryptographic signature to help prevent any MITM attacks. Additionally, attackers may conduct a denial-of-service attack (DoS) to make it difficult for authorised users to access the service by raising traffic or performing fraudulent transactions. However, because blockchain is decentralised, our model becomes resistant to DoS and DDoS-related attacks. Due to Ethereum's enormous number of mining nodes, it is extremely resistant to DDoS attacks.

\section{Conclusions}
This paper presented the design and implementation of a blockchain-based solution for digital vaccination and testing certificate. The proposed approach utilises the Ethereum blockchain system and smart contract technologies to facilitate secure vaccination certificate creation and verification.  Additionally, our system will feature a decentralised identity architecture that will be responsible for managing users' identities. This will be accomplished by using self-sovereign identification (SSI) reinforced by solid security design standards included on an Ethereum blockchain smart contract. Finally, the security of our system and an attacks model have been analysed to prove the feasibility of our system. It is feasible that our approach satisfies the security requirements for vaccination certificate and meet future demands. Our solution showed enhancements in security and users' privacy. In addition, we provided an analysis of the performance and the associated transactions costs. We observe that our approach provides negligible memory and CPU usage. However, the transactions cost and key management are the major drawbacks that need to be considered. Finally, we hope that our design provides advantages in the area of user authentication and a source of motivation for further research into this field.

\vspace{12pt}

\bibliographystyle{IEEEtran}  
\bibliography{paper} 
\end{document}